\documentclass[mathleft
]{an}
\usepackage{graphicx}
\usepackage{times}
\usepackage{psfig}
\usepackage{natbib}
\usepackage{amsmath}
\usepackage{amssymb}
\newcommand{\kms}{km s$^{-1}$}
\newcommand{\ha}{H$\alpha$}
\newcommand{\hb}{H$\beta$}

\newcommand{\ebv}{E(B-V)}
\newcommand{\lala}{$\lambda\lambda$}

\begin{document}

\Pagespan{789}{}
\Yearpublication{2008}%
\Yearsubmission{2008}%
\Month{11}%
\Volume{999}%
\Issue{88}%

\title{The host galaxies of Compact Steep Spectrum and
  Gigahertz-Peaked Spectrum radio sources
        }

\author{J. Holt\inst{1}\fnmsep\thanks{Corresponding author:
  \email{jholt@strw.leidenuniv.nl}\newline}
}
\titlerunning{Host galaxies of CSS \& GPS sources}
\authorrunning{J. Holt}
\institute{Leiden Observatory, Leiden University, PO Box 9513, 2300 RA
  Leiden, The Netherlands. 
}

\received{2008}
\accepted{}
\publonline{}


\abstract{%
I will review some of the developments in studies of the host galaxy
properties of Compact Steep Spectrum (CSS) and GigaHertz-Peaked
Spectrum (GPS) radio sources. In contrast to previous reviews
structured around observational technique, I will discuss the host
galaxy properties in terms of  morphology, stellar content and
warm gas properties and discuss how compact, young radio-loud AGN
are key objects for understanding galaxy evolution.
}

\maketitle

\section{Introduction}
In recent years it has become increasingly clear that AGN play a key
role in galaxy evolution, in particular the way in which AGN and its host
galaxy interact (e.g. Silk \& Rees 1998; Fabian 1999; di Matteo
et al 2005).  A key time to study the impact the AGN
 will have on its host galaxy, and how the
properties of the host galaxy will influence the  AGN, is during the
early stages of evolution. As discussed in the papers listed above,
among others, 
after relocating to the centre of the galaxy after the merger, the central
black hole will grow rapidly through merger-induced accretion,
eventually `switching-on' and becoming a quasar. The merger will
deposit large quantities of gas and dust into the nuclear regions,
essential to fuel the young AGN, and this must eventually be shed
through outflows and winds. However, much of this initial,
rapid growth phase will be  obscured from view at optical wavelengths
by the dense natal cocoon.

Cue the Compact Steep Spectrum (CSS) and GigaHertz-Peaked Spectrum
(GPS) radio sources. Now believed to be small due to
evolutionary stage, rather than confined and frustrated old sources
(e.g. Fanti et al. 1990; Fanti et al. 1995; Owsianik et al. 1998;
Murgia et al. 1999), the compact radio sources
provide a unique tool for 
pinpointing young, recently  triggered AGN.

Until recently, most effort has focused on the AGN and it's radio
jets, particularly in the radio band. However, the host galaxies can
provide a number of key results, important for both understanding
compact radio sources themselves and how they fit in to the bigger
picture. Through deep optical spectroscopy with 4-m and 8-m class
telescopes, it is possible to study both the emission lines and the 
near UV-optical continuum emission to search for signatures of
AGN-host galaxy interaction and probe the stellar populations \& star
formation history. 

In these proceedings, I give a brief overview of some of the recent
developments in studies of compact radio source host galaxy
properties. Instead of focussing on observation wavebands and
techniques, we now know enough to discuss the various properties in
terms of physically meaningful groups, namely the overall system
morphology, the stellar and gas content and evidence for
AGN-feedback.

\section{Host morphology}
\label{morphology}
\begin{figure*}
\centerline{\psfig{file=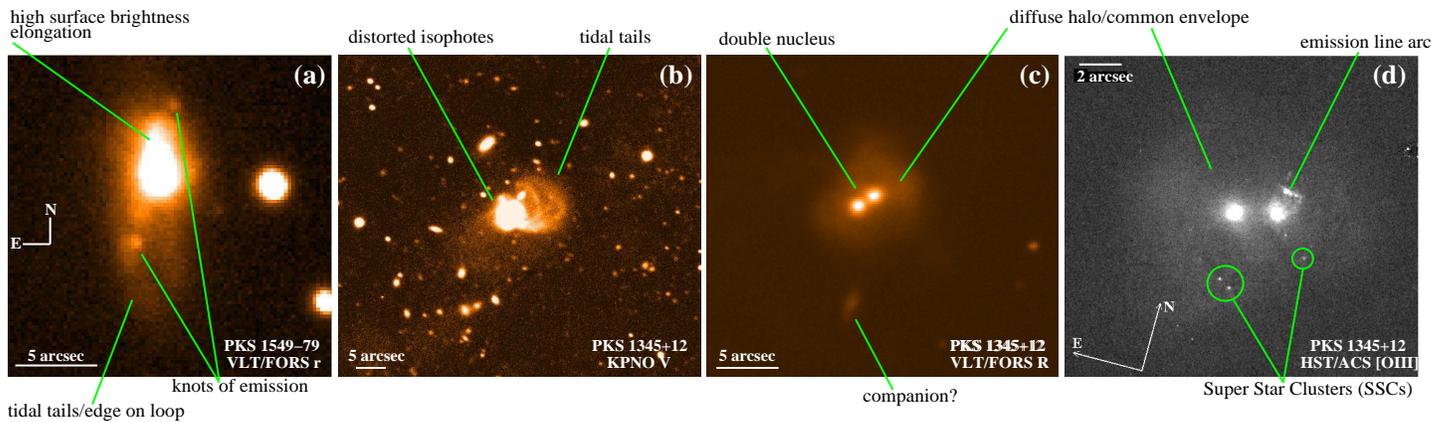,width=19cm,angle=0.}}
\caption[Morphologies of PKS 1345+12 and PKS 1549-79.]{(a)
VLT/FORS1 Gunn r image of the southern compact flat spectrum radio
source PKS 1549-79 taken from Holt et al. (2006). The image clearly
highlights a disturbed morphology with tidal tail like features
extending south from the nucleus and several knots of star formation
in the diffuse halo are clearly seen. (b), (c) \& (d) images of the
GPS source PKS 1345+12. The largest scale image (b) is a Kitt Peak V
band image from Emonts (priv. comm.) showing large-scale tidal-tail
like features to the NW. A VLT/FORS2 R band image reveals the double
nucleus in a common envelope with what appears to be a companion?
galaxy to the S. HST/ACS with a narrow band filter tuned to {[O III]}
emission reveals structure in the common envelope including an
emission line arc to the NW of the W nucleus and several knots of star
formation in the halo (see also Rodriguez Zaurin et al. 2007). These four
images suggest that PKS 1549-79 and PKS 1345+12 have both undergone
major mergers in their recent past, which triggered both the nuclear
activity and star formation in the host galaxies.}
\label{fig:morphology}
\end{figure*}
One of the most basic and easily determined  properties of the host
galaxies is morphology. Over the last 20 years, a number of broad band
imaging studies have been carried out (e.g. Gelderman \& Whittle, 1994; O'Dea
et al. 1996 \& references within). From these studies it is clear that
{\it all} CSS hosts (Gelderman \& Whittle 1994), and {\it the majority}
($\sim$60\%) of GPS hosts (O'Dea et al. 1996) show evidence 
for recent interactions and/or mergers (e.g. tidal tails, double nuclei,
distorted isophotes and arcs, shells and knots of emission) in which at least
one of the galaxies involved is gas-rich (Stanghellini 1993).
 What appear to be companion galaxies are also commonly
observed.
The results for compact radio sources are consistent to those found for
studies of powerful  radio galaxies (e.g. Heckman et al. 1986; Smith \&
Heckman 1989), in particular the hosts of FR II radio sources (O'Dea
1998).  Figure {\ref{fig:morphology}} shows some recent optical
images of two compact radio sources.

In the past, much effort has gone
into classifying hosts as 'galaxies' or 'quasars' and treating them as
discrete samples. This classification should be used with
caution for two reasons. First, if we believe in unification, all
compact (young) radio sources should be similar objects and the host galaxy
classification will only depend on viewing angle\footnote{Note, some
  extended radio sources may appear compact due to projection effects
  but should only account for $\lesssim$25-30\% of the 
  identified compact radio sources (Fanti et al. 1990)}. Second,  as
discussed above, the majority of CSS/GPS sources 
show evidence for a recent merger or interaction which will have injected
a significant amount of gas and dust into the circumnuclear
regions. This  will obscure the young, recently triggered AGN during
the early stages of it's evolution (e.g. Silk \& Rees 1998; Fabian
1999). Given this, it is likely that we may not be able to see directly into the
nucleus in compact (young; e.g. Fanti et al. 1990) radio sources and
could mistakenly classify a young `quasar' as a 
`galaxy-type' host (e.g. PKS 1549-79; Holt et al. 2006).

\section{Stellar content}
\label{stars}
\begin{figure*}
\centerline{\psfig{file=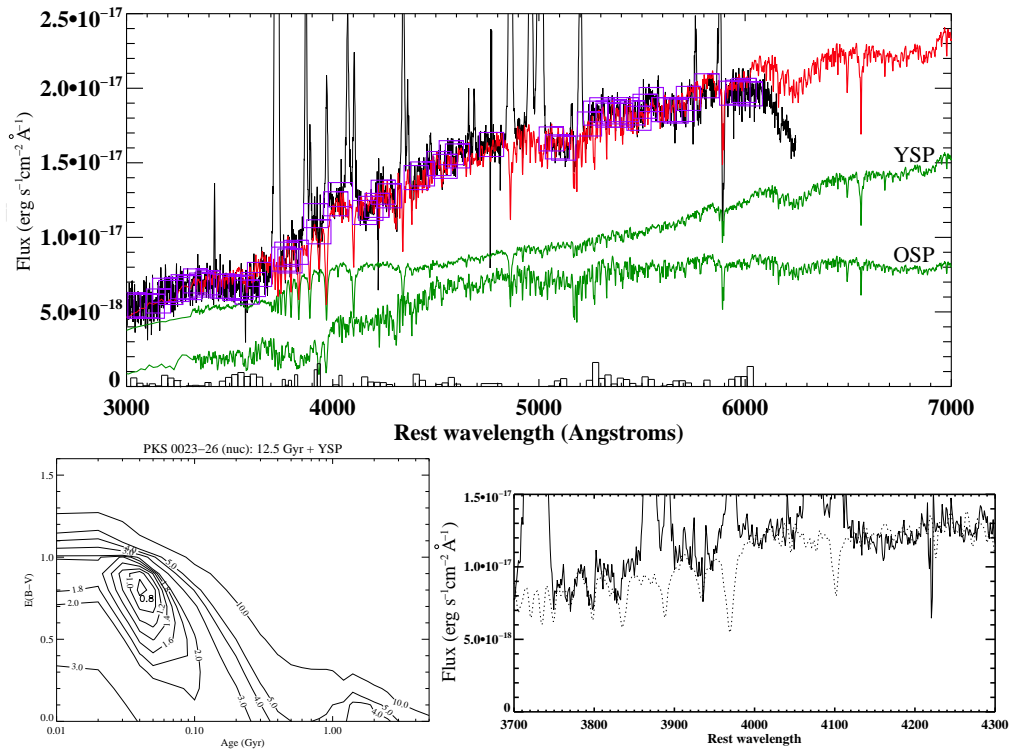,width=13.5cm,angle=0.}}
\caption[PKS 0023-26 SED.]{Results of SED modelling to the nucleus of
  the southern CSS source PKS 0023-26 taken from Holt et
  al. (2007). {\it Top plot}: The black line traces the rest frame
  nuclear spectrum which has been  corrected for Galactic
  extinction and the nebular continuum (which
  accounts for 22.5\% of the flux at 3540-3640\AA) has been subtracted. Overplotted is the
  best fitting model (red line) comprising a 12.5 Gyr OSP and a 0.03
  Gyr YSP, reddened by a foreground screen with E(B-V) = 0.9. The
  purple boxes show the location of the bins used for fitting and the
  histogram on the $x$-axis shows the residuals of the fit by bin. {\it
    Bottom left}: Contour plot of the reduced chi-squared
  ({$\chi_{\rm red}^{2}$}) space for combinations of YSP age \&
  reddening -- the minimum highlights the better fits. {\it Bottom
    right}: Once the best fits are identified using the whole SED and
  $\chi_{\rm red}^{2}$ contour plot,
  detailed comparisons are made between the data (solid line) and models
  (dotted line) using the stellar absorption features,
  particularly Ca H+K and the Balmer lines.
}
\label{fig:sed}
\end{figure*}
To date, the only systematic studies of the stellar content of the
host galaxies of compact radio sources have used optical and/or
near-IR imaging (among others: Snellen et al. 1996a,b, 1998; O'Dea et
al. 1996; de Vries et al. 1998, 2000, de Vries 2003). Using colours to estimate the
broad-band SEDs (R,J,H,K), these studies find the colours of both
CSS and GPS sources to be consistent with old stellar populations
(OSPs; \newline z$_{formation} \gtrsim$ 5), typical of passively or non-evolving 
elliptical galaxies. Similar results were found in the comparison
samples of FR IIs. 

Given the large number of CSS and GPS sources displaying evidence for
interactions (see Section \ref{morphology}), and knowing that gas-rich
mergers can also trigger starbursts, it is surprising that
evidence for this was not observed in the stellar populations. This
mis-match in results was noted by 
 de Vries (2003) but was explained as observing `the first of many
interactions'. 

However, it has been known for sometime that, when bluer rest-frame
colours are considered, many radio galaxies show ultraviolet (UV) excesses
compared to normal, passively evolving elliptical galaxies (e.g. Lilly
\& Longair 1984; Smith \& Heckman 1989). The active nucleus
will clearly contribute to the UV light, through one or more processes
(e.g. direct and/or scattered light from the AGN, emission lines and
nebular continuum; e.g. Tadhunter et al. 2002 and references
within). However, recent studies have shown that, by taking particular
care to account for all of the activity-related components, it is
possible to use spectral synthesis modelling techniques across the
entire optical rest-frame SED to reveal the presence of young stellar
populations (YSPs) in radio galaxies (e.g. Holt et al. 2007 and references
within). 

As part of the above studies, young and intermediate age stellar
populations have been found  in
8 compact radio sources including both CSS and GPS sources (PKS 
0023-26, PKS 2135-209, PKS 1345+12, 3C 213.1  \& 9C J1503+4528), a
compact flat spectrum source (PKS 1549-79) and extended radio sources
with a second, smaller-scale compact core radio source  (3C 236 \& 3C 459). 
The key results are:
\begin{itemize}
\item {\it Young stellar populations: few Myr - 1 Gyr}. Some sources
  are clearly in the first throes of activity (very young YSP) whilst
  others may be re-triggered AGN? (older YSP); Wills et
  al. (2008). The YSPs are often significantly reddened (Holt et al. 2007).
\item {\it The starbursts are  galaxy wide events}. In sources which
  are spatially resolved, the UV excess is extended (Holt et al. 2007;
  see also HST UV imaging by Labiano et al. 2008). In some sources,
  extended apertures have been 
  modelled showing clear evidence for YSPs (Tadhunter et al. 2005)
  including very young Super 
  Star Clusters (SSCs) in the haloes of some objects (Rodriguez Zaurin et al. 2007). 
\item {\it Prolific star formation}. Overall, the YSPs account for few-100\% of
  the stellar mass (Holt et al. 2007). Some of the compact radio sources are also
  classified as ULIRGs.
\end{itemize}

\section{Gas content}
\label{gas}
Another major component of the host galaxies is gas and dust. As
discussed above, the majority of CSS/GPS sources show evidence for 
recent mergers/interactions in which at least one of the galaxies was
gas rich. Hence, large quantities of gas and dust will be deposited in
the circumnuclear regions. The gas may be split into different phases
(cold/warm/hot). In this review, I will focus on the warm (optical
emission line) gas, discussing the kinematic properties, physical
conditions and dominant ionisation mechanism(s). For discussions on
the cold and hot gas, readers should refer to other contributions in
these proceedings (e.g. Fanti, Orienti, Morganti, Siemiginowska).

\subsection{Outflows}
\label{outflows}
\begin{figure*}
\centerline{\psfig{file=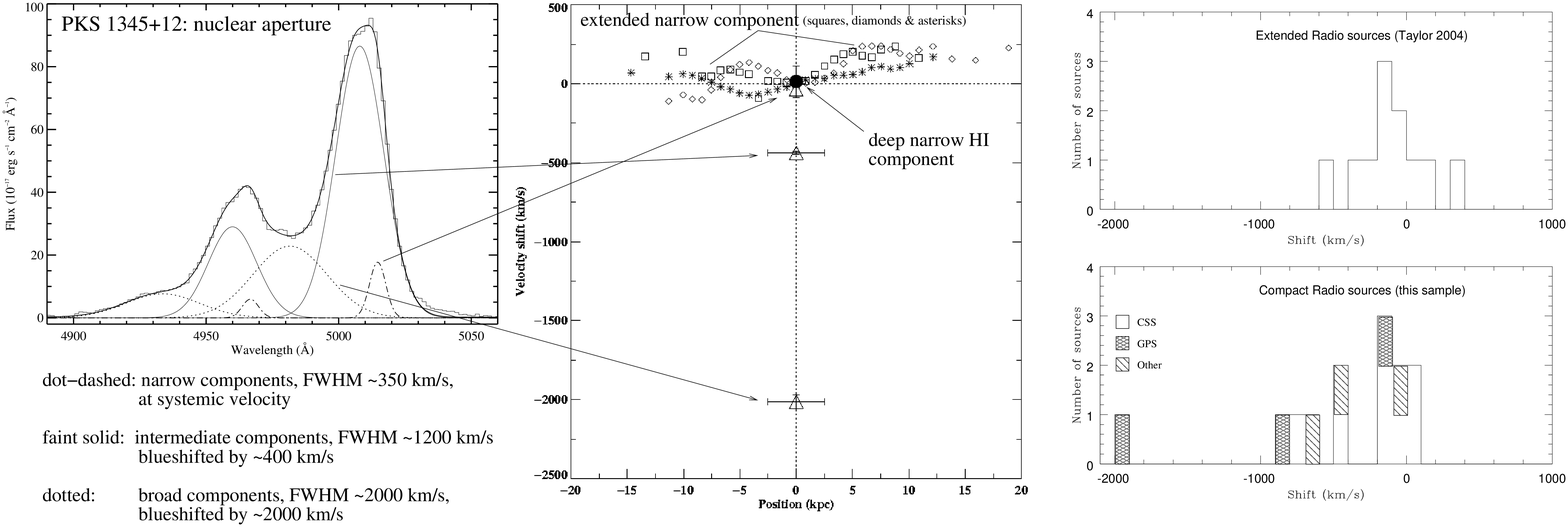,width=18cm,angle=0.}}
\caption[PKS 1345+12: {[O III]} and radial velocity profile.]{{\it
    Left}: {[O III]}\lala4959,5007 emission line profile in the nucleus of the GPS
  source PKS 1345+12 taken from Holt et al. (2003). The broad, highly
  complex profile of the doublet requires 3 Gaussian components to
  each line. The components are overplotted onto the {\it middle
    panel} which is a radial velocity profile of the system. 
The narrow component is consistent with the extended
  narrow component and therefore the systemic velocity of the
  system. Hence, the broader components are {\it blueshifted} with
  respect to the systemic velocity. {\it Right}: Histograms showing
  the narrow-broadest component velocity shifts for the whole sample
  of compact radio sources (14 sources; bottom panel) of Holt et
  al. (2008) and a comparison sample of nuclear apertures in extended radio sources (top
  panel). More extreme outflow velocities are observed in the nuclear
  apertures of compact  radio sources compared to extended radio
  sources. This trend continues within the compact radio source sample
  with the most extreme outflow velocities observed in the GPS
  (i.e. {\it smallest} sources.
}
\label{fig:outflows}
\end{figure*}
In 1994, Gelderman \& Whittle reported that the nuclear emission lines
in a sample of 20 CSS sources were strong, with broad and structured {[O
    III]}$\lambda$5007 profiles which may be consistent with strong
interactions between 
the radio source and the ambient gas. However, it was not until 2001
that unambiguous evidence for outflows in the emission line gas were
found in the southern compact radio source PKS 1549-79 (Tadhunter et
al. 2001). 

Tadhunter et al. (2001) reported that in PKS 1549-79, the high
ionisation lines (e.g. {[O III]}\lala4959,5007) were both significantly broader
(FWHM$_{[O III]}$ $\sim$ 1350 \kms~ compared to FWHM$_{[O
    II]}$ $\sim$ 640 \kms) and blueshifted ($v \sim$ 600 \kms) with
respect to the low ionisation lines (e.g. {[O II]}\lala3727). Combined with
their other data, they interpreted this as the spectral signature of
an outflow in the emission line gas, driven by the young, small,
expanding radio jets (see Figure 2 in Tadhunter et al. 2001). 

The discovery of the outflow in PKS 1549-79 triggered a detailed study
of the  emission line gas in a sample of 14 compact radio
sources (Holt 2005; Holt et al. 2008a). Careful modelling of the
narrow extended line emission allows the systemic velocity to be accurately
measured (Figure \ref{fig:outflows}). In  the nuclear apertures, the
lines are significantly 
broader and often require multiple Gaussian components. Combined with
the accurate systemic redshifts, it is clear that in 11/14 sources, the broader
components are significantly {\it blueshifted}, tracing fast
outflows\footnote{The blueshifted components are thought to be
  outflows rather than inflows using
  reddening arguments. See Holt et al. 2003 for a detailed argument.}
in the emission line gas (Holt et al. 2008a). The most
extreme example is observed in 
the GPS source PKS 1345+12 ($v \sim$ 2000 \kms; see Figure
\ref{fig:outflows} and Holt et al. 2003). O'Dea et al. (2002) also report fast
 ($v \sim$ 300-500 \kms) outflows in 3 CSS sources from HST/STIS
spectroscopy, 2 of which are also in the sample of Holt et al.

Whilst the samples are small, it is clear that the nuclear kinematics
in compact radio sources are more extreme than in extended radio
sources (see Figure \ref{fig:outflows} and Holt et al. 2008a). This
trend extends within the sample -- the smallest (GPS) radio sources
tend to have the most extreme outflow velocities. However, radio
source orientation may also play a role (Holt et al. 2008a). 

Whilst the expanding radio jets provide a convenient driving mechanism
for the observed nuclear outflows, it is important to test this
scenario. Using emission line ratios and diagnostic diagrams, it is
possible to establish the dominant ionisation mechanism of the
gas. For their sample of 14 compact radio sources, Holt (2005) \& Holt
et al. (2008b) found that, whilst the nuclear narrow components (at the systemic
velocity) were typically photoionised by the AGN, the broader,
blueshifted components were more consistent with shock ionisation,
with fast velocities ($v_{shock}$ $>$ 300 \kms) and a pre-photoionised (precursor)
component. Similar results were found by Labiano et al. (2005) for
their sample of 3 CSS sources with HST/STIS (2 of which are in the
Holt et al. sample).
\begin{figure*}
\centerline{\psfig{file=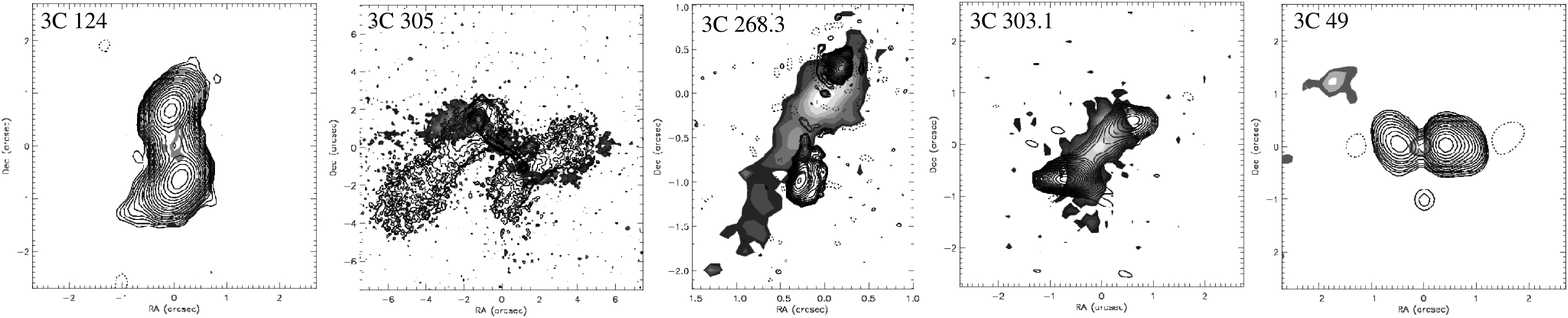,width=18cm,angle=0.}}
\caption[Radio-optical alignments.]{Examples of radio-optical emission
  alignments in compact radio sources taken from Privon et
  al. (2008). The greyscale traces the optical line emission and
  the contours represent the radio emission. See Privon et al. (2008)
  for details.
}
\label{fig:ro}
\end{figure*}

\begin{figure}
\centerline{\psfig{file=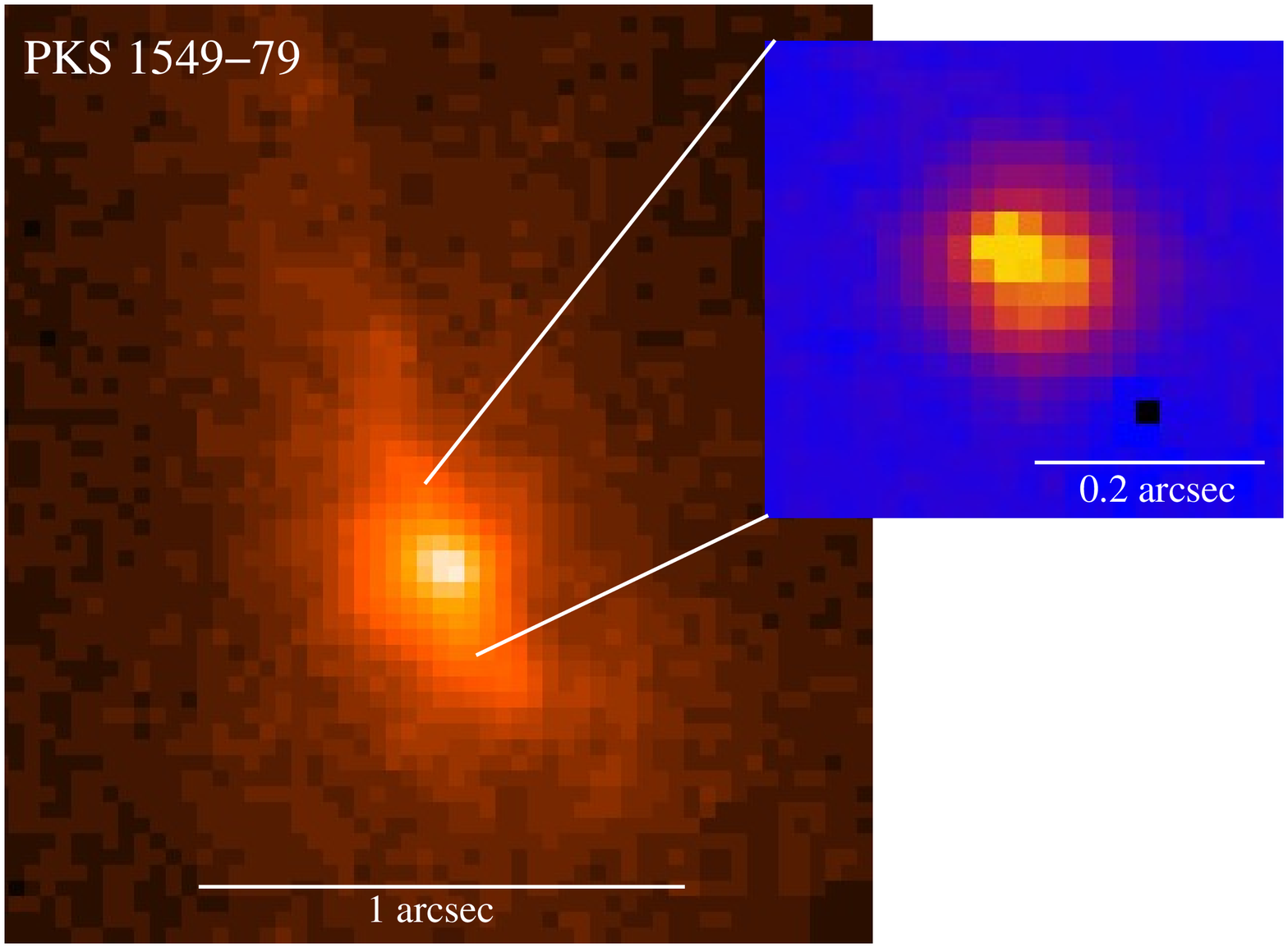,width=8cm,angle=0.}}
\caption[Radio-optical alignments.]{HST/ACS broad and narrow band ({[O
      III]}) imaging of the compact flat-spectrum radio source PKS
  1549-79 (see Batcheldor et al. 2007). The zoom shows the emission
  line outflow in this source is 
  spatially resolved, ruling out large-scale starburst driven
  winds. Whilst there is tentative suggestion that the line emission
  is elongated in the direction of the radio source, at the moment, it
  is not possible to distinguish between AGN-winds and radio-jet
  interactions as the dominant outflow driving mechanism. 
}
\label{fig:1549}
\end{figure}

A key advantage of emission line outflows over absorption line
outflows is that, in addition to kinematics etc from spectroscopy, the
outflowing regions can be directly imaged to reveal the size, location
and morphology. Narrow band HST imaging (e.g. {[O II]} and/or {[O
    III]}) has shown that, in CSS sources, the optical line emission
is co-spatial with, and strongly aligned with, the radio source at {\it
  all redshifts} (e.g. de Vries et al. 1997,1999; Axon et al. 2000;
Privon et al. 2008) -- see Figure {\ref{fig:ro}}. This is in contrast
to extended radio sources in which the alignment is only observed at z
$>$ 0.6. Batcheldor et
al. (2007) have also attempted to 
image the emission line outflow in the GPS source PKS 1345+12 and the
compact flat spectrum radio source PKS 1549-79 (see Figure
\ref{fig:1549}). Whilst the radio and
optical line emission are clearly co-spatial, the scales are too small
to give more than a tentative suggestion that the line emission is
elongated along the radio axis.  Hence, whilst in the larger CSS sources,
there is a clear argument for jet-driven outflows but in the smaller
GPS sources, it is impossible to distinguish between AGN-winds and
jet-driven outflows.

\subsection{Physical conditions}
It is clear that, in order to trigger and fuel nuclear activity, large amounts
of gas and dust must be deposited into the circumnuclear regions,
whatever the triggering mechanism. For
compact radio sources, an ongoing debate has been whether there is
sufficient material in the nuclear regions to confine and frustrate
the radio source or whether small scale radio jets indeed represent an
early evolutionary stage. It is therefore important to determine the
physical conditions of the gas, in particular the gas density and
mass.

Using emission line ratios, we can make the following general
statements about the circumnuclear gas in compact radio sources:
\begin{itemize}
\item {\it Significant reddening in some sources, not in
  others}. Reddening can be measured using the Balmer lines \newline
  (\ha,\hb~etc). In some sources, large reddening is measured (up to
  \ebv~= 2.0 in the NLR) which can increase with component FWHM in a particular
  source, as in PKS 1345+12 (Holt et al. 2003; Holt et
  al. 2008b). In general, the reddening measurements are not
  significantly different to those for extended radio sources
  (e.g. Morganti et al. 1997; Labiano et al. 2005; Holt et
  al. 2008b). The measurements of large reddening, particularly in the
  broader components, has strengthened the interpretation of the
  blueshifted line emission as outflows rather than inflows. Whilst
  the NLR typically shows evidence for low-moderate extinction, the
  nucleus itself can still be much more highly extinguished at optical
  wavelengths (e.g. PKS 1549-79; Holt et al. 2006).
\item {\it Large gas densities}. The electron densities are high,
  typically n$_{e}$ $>$ ~few 1000 cm$^{-3}$ (Holt et
  al. 2003,2008b). Unfortunately, 
  the line profiles are highly complex, precluding accurate
  measurements using the traditional {[S
      II]}$\lambda$6716/$\lambda$6731 ratio. Work is currently
  underway to measure the gas densities using alternative
  diagnostics (Holt et al. 2009). 
\item {\it High gas temperatures}. Electron temperatures are estimated
  to be T$_{e}$ $\sim$ few $\times$ 10$^{4}$ K, although line
  measurements suffer from similar issues to those discussed above
  (Holt 2005; Holt et al 2008b).
\item {\it Mass of the wam gas}. Warm gas masses  can be estimated
  using the luminosity of \hb~and the estimated electron densities. To
  date, this has only been estimated for the GPS source PKS 1345+12 as
  $M_{gas}$ $<$ 10$^{6}$ M$_{\odot}$ for the line emitting gas. 
\end{itemize}

\section{The bigger picture}
\label{bigpicture}
The results above, combined with previous work, are consistent with
the following picture:
\begin{enumerate}
\item A major merger occurs where at least one galaxy is gas rich
  (c.f. observed morphologies). This merger deposits a dense and dusty
  natal coccoon (c.f. large measured densities and reddening along
  with the compact radio source \& proto-quasar PKS 1549-79 in which
  the quasar is totally obscured at optical wavelengths). 
\item Early on in the merger, a large burst of star formation is
  triggered  in the host galaxy  (c.f. detection of YSPs in several
  compact radio sources).
\item At some point after the onset of the merger, the nuclear
  activity \& radio source are triggered (c.f. age of the YSPs and the
  age of the radio source from kinematical and synchrotron spectral
  ageing measurements; $t_{radio} <$ 10$^{4}$ years). 
\item As the radio source evolves, the small-scale jets expand through
  the natal cocoon, driving outflows in the
  emission line gas (c.f. fast outflows, high temperature and evidence
  for jet-cloud interactions/shock-ionisation). 
\item Eventually the AGN will shed its coccon, starving the nucleus
  and the circumnuclear starburst (c.f. we observe `naked quasars'
  and there is no current evidence to suggest there is sufficient gas
  in the nuclear regions to confine and frustrate the radio source). 
\item In at least some sources, after a period of quiescence, nuclear
  activity is restarted (e.g. double-double sources and the discovery
  of faint, diffuse, extended radio emission around a number of
  compact radio sources (e.g. Stanghellini et al. 2005). 
\end{enumerate}

With the observations to date, we have been able to form an
evolutionary scenario which fits in well with galaxy evolution in
general. However, there are still some uncertainties which need to be
addressed. 
Whilst morphology/merger evidence is observed in all CSS
  sources, it is not observed in all GPS sources ($\sim$60\%~show
  evidence for morphological disturbance). This leads us to question
  whether this is due to previous data quality (many of the GPS
  imaging results are at least 10 years old and may be improved
  with larger telescopes) or whether an alternative triggering
  mechanism is required for some GPS sources. Are these sources
  undergoing recurrent activity? Or a different accretion mode? -- see
  e.g. Tasse (2008), 
Best et al. (2005) for discussions on hot/cold accretion modes for
`radio-mode' and `quasar-mode' AGN. 

The stellar population studies on radio galaxies, including a
  few CSS/GPS sources, have now brought into question previous work
  suggesting the host galaxies of CSS/GPS sources were passively or
  non-evolving elliptical galaxies. The past studies included objects
  which have now been confirmed to have large, young stellar
  populations (e.g. PKS 1345+12). However, the number of CSS/GPS
  sources investigated so far is small. 

The majority  of the CSS/GPS sources included in the studies
  discussed above are powerful radio sources. However, large numbers
  of compact radio sources are significantly weaker at radio
  wavelengths. It is not known how these sources compare to the new
  results on powerful sources. 

Finally, in  galaxy evolution simulations, the models require a
  significant amount of energy for AGN feedback -- 5-10\% of the
  accretion energy is required. Fast outflows are now observed in many
  CSS/GPS sources in both the warm (optical emission lines) and cold
  (radio; see Morganti et al., these proceedings) gas. However,
  current estimates of the mass outflow rates  for both phases of the
  ISM fall well below that required for galaxy evolution simulations
  (e.g. PKS 1549-79; IC 5063, Morganti et al. 2007). Work is currently
  underway to better estimate the mass outflows rates in compact radio
  sources. 

Hence, much work still needs to be done, especially as compact radio
sources are ideal probes of early AGN evolution. 
More detailed studies are required to enable us to disentangle the interplay
between the active nucleus and the host galaxy.

\acknowledgements
JH acknowledges a PPARC PDRA \&  and a NWO Post-doc position.


\end{document}